\documentclass[12pt]{article}

\usepackage{graphicx}
\usepackage{bm}

\begin{document}


\centerline{\Huge\bf Quantum time ordering and degeneracy}
\vskip 1em
\centerline{\LARGE \bf II: Coherent population transfer }
\centerline{\LARGE \bf between degenerate states}

\vskip 2em

\vskip 2cm

\centerline{\it Kh.Yu. Rakhimov$^{1,2}$, Kh.Kh. Shakov$^1$, and J.H. McGuire$^1$}
\vskip 1cm
\centerline{ $^1$Department of Physics, Tulane University, New Orleans, LA 70118, USA}
\centerline{ $^2$Department of Heat Physics, Uzbekistan Academy of Sciences,}
\centerline{ 28 Katartal St., Tashkent 700135, Uzbekistan}

\date{\today}

\begin{abstract}
We find conditions required to achieve complete population transfer,
via coherent population trapping, from an initial state to a designated
final state at a designated time in a degenerate $n$-state atom, where
transitions are caused by an external interaction.
In systems with degenerate states there is no time ordering.
Analytic expressions have been found for transition
probabilities in a degenerate $n$-state atom interacting
with a strong external field that gives a common time
dependence to all of the transition matrix elements.
Except for solving a simple $n^{th}$ order equation to
determine eigenvalues of dressed states, the method is
entirely analytic.  These expressions may be used to
control electron populations in degenerate $n$-state atoms.
Examples are given for $n = 2$ and $n=3$.
\end{abstract}

\newpage

\section{Introduction}

Analytic descriptions of physical systems are convenient for understanding
quickly and easily how a system works under various conditions.
An example is population control in quantum systems, namely transfer of electrons
from an ensemble of atoms all in the same initial state to specified
final states within the ensemble.
This is used in problems ranging from quantum information \cite{nc00,bez00}
to chemical dynamics \cite{Rabitz93,kas02}.
Such problems are often studied in terms of a single
electron in an atom with $n$ discrete states interacting with a
strong external field \cite{shore,me,meystre,ae,gue97,kl87,bb77}.
In this paper analytic expressions are presented for the transition probabilities
as a function of time in a degenerate $n$-state atom.
Once analytic solutions are obtained, corrections due to finite energy
splittings can be introduced, e.g. in numerical calculations.
Then real systems with finite differences in the energies of the $n$ states
can be studied.

Our degenerate energy approximation is somewhat similar to the
rotating wave approximation (RWA) \cite{shore,me} that
has been widely applied to both $2$ and $3$-state atomic models.
In RWA, however, degenerate atomic states are not used.
Instead one tunes the frequency of the external field $V(t)=V_0\cos(\omega t)$
to the frequency difference of two non-degenerate states
so that the detuning parameter $\Delta = \hbar (\omega - \omega_{12})$
tends to zero, where $\omega_{12}=(E_1-E_2)/\hbar$.
Thus in RWA an initial state of an atom plus one
photon is degenerate in energy with the final state of the atom.
An advantage of using degenerate atomic states is that one is not restricted
to external interactions with frequencies close to the transition frequency.
Thus $\omega$ can be used for control, e.g. to vary the duration of time
that the transferred population remains in the designated state, or to reduce
the population leakage that occurs when the energy states are not fully degenerate.
On the other hand RWA has the advantage that analytic solutions
have been found when the detuning parameter is small, but finite.

\section{Theory}

Let us consider an $n$-state atom interacting with an external field, $V_{ext}(\vec{r},t)$.
The total Hamiltonian for this system is $H = H_0 + V_{ext}(\vec{r},t)$.  The $n$ eigenstates,
$\phi_k$, and corresponding eigen-energies, $E_k$, of $H_0$ are assumed to be known.
The total wavefunction may be expanded in terms of the known eigenstates, namely,
$\Psi(t) = a_1(t) \phi_1 + a_2(t) \phi_2 + \cdots + a_n(t) \phi_n$.
With atomic units, using $i \dot{\Psi} = (H_0 + V_{ext}(\vec{r},t)) \Psi$,
with $H_0 \phi_k  = E_k \phi_k$ and $\int \phi^*_j \phi_k d\vec{r} = \delta_{jk}$,
one then obtains \cite{me},
\begin{eqnarray}
\label{nexact}
i \dot{a}_j(t) = E_j a_j(t) + \sum_{k=1}^n V_{jk}(t) a_k(t)  \ \ ,
\end{eqnarray}
where $V_{jk}(t) = \int \phi^*_j V_{ext}(\vec{r},t) \phi_k d \vec{r}$.
These equations are exact for an $n$-state atom.

We assume that the system is degenerate, namely that
all the energies, $E_j$, are the same.
Since the zero point of energy is arbitrary, one may generally set $E_j = 0$.
These conditions give the coupled equations for our degenerate $n$-state system, namely,
\begin{eqnarray}
\label{ndegen}
i \dot{a}_j(t) =  \sum_{k =1}^n V_{jk}(t) a_k(t)  \ \ .
\end{eqnarray}
We additionally require that all of the $V_{jk}(t)$ have the same time dependence,
e.g. $V_{jk}(t) = r_{jk} E_0 f(t)$, where $f(t)$ is a real, but otherwise
arbitrary function of time.
For this work we consider $V_{jk} = V_{kj}$ to be real.
We use the initial conditions $a_1(0) = 1$, and $a_j(0) = 0$ for $j \neq 1$.

\subsection{2-state atom}

We first illustrate our scheme for population control in a two state atom.
For $n = 2$ Eq(\ref{ndegen}) becomes,
\begin{eqnarray}
\label{2degen}
i \dot{a}_1(t) &=&  V_{11}(t) a_1(t) + V_{12}(t) a_2(t) \nonumber \\
i \dot{a}_2(t) &=&  V_{21}(t) a_1(t) + V_{22}(t) a_2(t) \ \ .
\end{eqnarray}
Set $V_{12}(t) = V_{21}(t) = V(t)$, and $V_{jj}(t) = \epsilon_j V(t)$.

Now seek a solution $c(t) = x_1 a_1(t) + x_2 a_2(t)$ such that
$i \dot{c}(t) = z V(t) \ c(t)$, where $z$ is to be determined.
Since $a_1(0) = 1$ and $a_2(0) = 0$, we may set $x_1 = 1$.
Using Eq(\ref{2degen}), one has,
\begin{eqnarray}
\label{2c}
  i \dot{c}(t) &=& i \dot{a}_1(t) +  x_2 \ i \dot{a}_2(t)  \nonumber \\
   &=& \epsilon_1 V(t)  a_1(t) + V(t) a_2(t)
	+ x_2 \ V(t) a_1(t) + x_2 \ \epsilon_2 V(t) a_2(t) \nonumber \\
	&=& (x_2 + \epsilon_1) V(t) [ a_1(t) + \frac{1 + x_2 \epsilon_2}{x_2 + \epsilon_1} a_2(t)]
	\nonumber \\
	&=& z \   V(t) c(t) \ \  .
\end{eqnarray}
This holds if and only if  $z = x_2 + \epsilon_1$
and $\frac{1 + \epsilon_2 x_2}{x_2 + \epsilon_1} = x_2$.
This yields two roots for $x_2$ and $z$, namely,
$x_{(1,2)2} = \frac{\epsilon_2 - \epsilon_1}{2}
\pm \sqrt{1 + (\frac{\epsilon_2 - \epsilon_1}{2})^2}$
and $z_{1,2} =  \frac{\epsilon_2 + \epsilon_1}{2}
\pm \sqrt{1 + (\frac{\epsilon_2 - \epsilon_1}{2})^2}$.
Using $x_{12} = x_{(1)2}$, $x_{22}= x_{(2)2}$, $x_{11} = x_{21} = 1$, and
defining the action, $A(t) = \int_0^t V(t') dt' $,
the corresponding eigenfunctions are,
\begin{eqnarray}
\label{2c-ans}
c_1(t) &=& e^{-i z_1 A(t)} =  x_{11} a_1(t) + x_{12} a_2(t) \nonumber \\
c_2(t) &=& e^{-i z_2 A(t)} =  x_{21} a_1(t) + x_{22} a_2(t)  \ \ .
\end{eqnarray}
The relation between the $c_i(t)$ and $a_j(t)$ may be expressed
as $c_i(t) = \sum_j^n {\cal M}_{ij} a_j(t)$, where ${\cal M}_{ij} = x_{ij}$.
This matrix may be inverted to give the population amplitudes,
\begin{eqnarray}
\label{2a-ans}
a_1(t) &=& \frac{1}{\Delta}[x_{22} c_1(t) - x_{12} c_2(t)]  \nonumber \\
a_2(t) &=& \frac{1}{\Delta}[ - x_{21} c_1(t) + x_{11} c_2(t)]  \ \ ,
\end{eqnarray}
where $\Delta = 2 \sqrt{1 + (\frac{\epsilon_2 - \epsilon_1}{2})^2}$
is the determinant of ${\cal M}_{ij}$.

A simple solution occurs when $\epsilon_1 = \epsilon_2 \equiv \epsilon$,
namely,
\begin{eqnarray}
\label{2a-ansa}
a_1(t) &=& e^{-i \epsilon A(t)} \ \frac{1}{2} [e^{-iA(t)} + e^{+iA(t)}]
	=  \ e^{-i \epsilon A(t)} \cos(A(t))  \nonumber \\
a_2(t) &=& e^{-i \epsilon A(t)} \ \frac{1}{2} [e^{-iA(t)} - e^{+iA(t)}]
	= - i e^{-i \epsilon A(t)} \sin(A(t))  \ \ .
\end{eqnarray}

One may now determine the conditions under which the population
of state 2 takes on any desired value $P_2 = P_v = v^2$, where $0 \leq v^2 \leq 1$.
Note that $P_1 = 1 - P_2$.  For Eq(\ref{2a-ansa}) this condition is satisfied if
$a_2(t) = \sin(A(t)) = v$ (within an overall phase), i.e., $A(t) = \sin^{-1}(v)$.
This can always be satisfied.  Note that for a given
$V(t)$ one may choose the time such that $A(t) = \sin^{-1}(v)$.
Alternatively to move population to this arbitrary value
at a particular time $t_0$, one may adjust $V(t)$ so that
$A(t_0) = \sin^{-1}(v)$.
If $\epsilon_1 \neq \epsilon_2$, it is easily shown that
$P_2(t) \leq 1/(1+ (\epsilon_2 - \epsilon_1)^2/4 ) < 1$.
Thus the diagonal matrix elements of $V_{ij}(t)$ prevent
complete transfer to the initially unoccupied state when they
are unequal, and simply contribute an overall phase when they are equal.

An interesting solution occurs when the ratio $ \chi / \omega = \pi / 2$, see~\cite{sm02}.
Then Eqs (\ref{2a-ansa}) take the form
\begin{eqnarray}\label{f8}
&& a_1=  \cos \lbrack  \frac{\pi}{2} \sin(\omega t) \rbrack \nonumber\\
&& a_2=i \sin \lbrack  \frac{\pi}{2} \sin(\omega t) \rbrack \ .
\end{eqnarray}
The corresponding occupation probabilities $P_1=|a_1|^2$ and $P_2=|a_2|^2$ are
\begin{eqnarray}\label{f9}
&& P_1=  \cos^2 \lbrack  \frac{\pi}{2} \sin(\omega t) \rbrack \nonumber\\
&& P_2=  \sin^2 \lbrack  \frac{\pi}{2} \sin(\omega t) \rbrack \ .
\end{eqnarray}
\begin{figure}
\centering
\scalebox{0.7}{\includegraphics{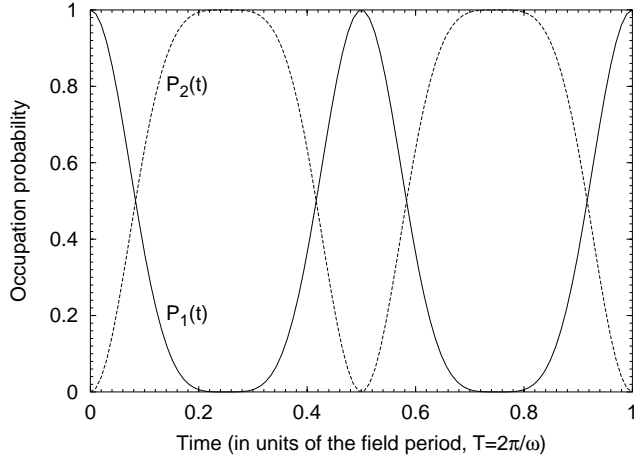}}
\caption{Analytical solution for $\chi/\omega=\pi/2$ for a system with two degenerate energy states.}
\label{fig1}
\end{figure}
For this special value of $\chi/\omega$ the probability is completely transfered back and forth between the two states with the period $\pi/\omega$ as shown in figure~\ref{fig1}.
It can be easily shown that when $P_2(t)$ reaches its maximum value at $t=\pi/2\omega$, a small deviation $\epsilon$ from the value $\pi/2$ in the ratio $\chi/\omega$ reduces $P_2$ from $1$ to $1-\epsilon^2$. This can be used to estimate how sensitive this approach is to fluctuations in the strength of the external field, $\chi$.
The period of oscillation of the occupation probabilities is two times smaller than the period of oscillation of the radiation field since $P_i=|a_i|^2$.

Complete population transfer occurs in general wherever the action, $A(t)=\int_0^t V_{21}(\tau) \ d \tau$, is an integer multiple of $h/4$.
Each cycle has a "flat" part where the probability value remains very close to the extreme values of 0 or 1 for an extended period of time (cf. figure \ref{fig1}).
As we can see, the external radiation field with the ratio $\chi / \omega=\pi / 2$ can be used to cause periodical population inversion of the electronic states (a physical realization of a quantum controllable system).

Equations (\ref{f9}) can be expanded in a power series in the vicinity of the point
$t_0=T/4$, where $P_2(t)$ reaches its first maximum. Then $P_2$ can be accurately
approximated by taking only first few terms in the series, which converges rapidly
for $\tau < T =2 \pi/\omega$. If one chooses the potential in the form
$V_{ext}(\vec r,t)=-\vec r\ \vec E(t)$,
the first, second and third derivatives of the function $P_2$ are zero at $t=t_0$, and
\begin{equation}\label{f12}
P_2(t_0+\tau)= 1+\frac{1}{4!} \frac{d^4 P_2}{d t^4}\tau^4 + {\cal O}(\tau^6)
\approx 1 - \frac{\pi^2}{16}(\omega \tau)^4 \ .
\end{equation}
Hence the occupation probability $P_2$ in the vicinity of its maximum can be approximated by a $4th$ degree polynomial.

\begin{figure}
\centering
\scalebox{0.7}{\includegraphics{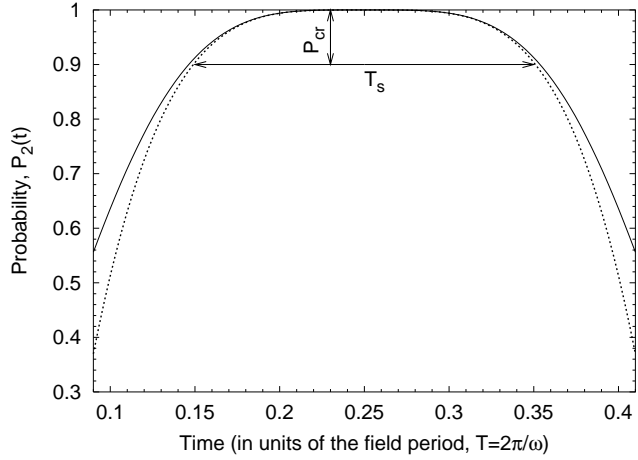}}
\caption{Population leakage in the vicinity of $t_0=T/4$. Solid line - analytic solution (\ref{f9}), dash-dotted line - polynomial approximation (\ref{f12}).}
\label{fig2}
\end{figure}

The polynomial approximation of Eq.(\ref{f12}) tells one how to choose the field frequency to provide a desired duration of the populated state. To obtain a duration $T_s$ of the populated state with a population leakage less than some critical value $P_{cr}$,
the field frequency to be used is
\begin{equation}\label{estim1}
\omega = \sqrt{\frac{4}{\pi}}\frac{(P_{cr})^{\frac{1}{4}}}{T_s}
\end{equation}
Near $\tau =0$ the error is ${\cal O}((\omega\tau)^6)$.
This is illustrated in figure \ref{fig2}.

\subsection{3-state atom}

Next consider a 3-state atom.
The general case for $n = 3$ can be solved by taking
$V(t) = \alpha^{-1} V_{12}(t) = \beta^{-1} V_{13}(t) = V_{23}(t) = \epsilon_j^{-1} V_{jj}(t)$.
Then Eq(\ref{ndegen}) becomes,
\begin{eqnarray}
\label{3degen}
i \dot{a}_1(t) &=&  \epsilon_1 V(t) a_1(t) + \alpha V(t) a_2(t) + \beta V(t) a_3(t) \nonumber \\
i \dot{a}_2(t) &=&  \alpha V(t) a_1(t) + \epsilon_2 V(t) a_2(t) + V(t) a_3(t) \nonumber \\
i \dot{a}_3(t) &=&  \beta V(t) a_1(t)  + V(t) a_2(t) + \epsilon_3 V(t) a_3(t)  \ \ .
\end{eqnarray}
Try $c(t) = x_1 a_1(t) + x_2 a_2(t) + x_3 a_3(t)$.  Again, $x_1 = 1$, and for convenience
we set $x_2 = x$ and $x_3 = y$.
Then,
\begin{eqnarray}
\label{3trial}
i \dot{c}(t) &=& i \dot{a}_1(t) + x \ i \dot{a}_2(t) + y \ i \dot{a}_3(t) \nonumber \\
 &=&  \  \epsilon_1 V(t) a_1(t) + \alpha V(t) a_2(t) + \beta V(t) a_3(t) \nonumber \\
	&& + x [\alpha V(t) a_1(t) + \epsilon_2 V(t) a_2(t) + V(t) a_3(t)] \nonumber \\
	&& + y [\beta V(t) a_1(t) + V(t) a_2(t) + \epsilon_3 V(t) a_3(t)] \nonumber \\
 &=& (\epsilon_1 + \alpha x + \beta y) V(t) \nonumber \\
 && \times[ a_1(t) + \frac{\alpha + \epsilon_2 x + y}{\epsilon_1 + \alpha x + \beta y} a_2(t)
  + \frac{\beta + x + \epsilon_3 y}{\epsilon_1 + \alpha x + \beta y} a_3(t) ] \ \ .
\end{eqnarray}
We seek $c(t)$ such that $ i \dot{c}(t) = z V(t) c(t)$.
The last line holds if and only if $z = \epsilon_1 +  \alpha x + \beta y$,
$ (\alpha + \epsilon_2 x + y)/(\epsilon_1 + \alpha x + \beta y) = x$
and $(\beta + x + \epsilon_3 y)/(\epsilon_1 + \alpha x + \beta y) = y$.
It can be easily shown that
$y = (\alpha (x^2 - 1) + (\epsilon_1 - \epsilon_2)x)/(1 - \beta x)$
and $x = (\beta(y^2 - 1) + (\epsilon_1 - \epsilon_3)y)/(1 - \alpha y)$.
After some algebra this leads to the cubic equation,
$ [(\alpha^2-\beta^2)+\alpha\beta(\epsilon_3-\epsilon_2)]x^3
+[\beta(2-\alpha^2-\beta^2)+\alpha(2\epsilon_1-\epsilon_2-\epsilon_3)
+\beta(\epsilon_1-\epsilon_2)(\epsilon_3-\epsilon_2)]x^2
+[(2\beta^2-\alpha^2-1)+\alpha\beta(2\epsilon_2-\epsilon_1-\epsilon_3)
+(\epsilon_1-\epsilon_2)(\epsilon_1-\epsilon_3)]x
+[\beta(\alpha^2-1)-\alpha(\epsilon_1-\epsilon_3)] = 0 $.
This yields three eigenvalues for $x$ (namely $x_j$) and three eigenvalues
for $y$ ($y_j$).

The eigenvalues, $\{x_j\}$ and $\{y_j\}$, determine eigenvalues for $\{z_j\}$,
and three eigenfunctions, $ c_j(t) = e^{-i z_j  A(t)}$, where
$A(t) = \int_0^t V(t')dt'$.  Specifically, $c_j=\sum_{i=1}^3{\cal M}_{ji}a_i$, where,
\begin{equation}
{\cal M}=\left(\begin{array}{ccc}
1&x_1&y_1\\
1&x_2&y_2\\
1&x_3&y_3\\
\end{array}\right)
\end{equation}
This matrix, ${\cal M}$, may be inverted, namely,
\begin{equation}
\label{Minv}
{\cal M}^{-1}=\frac{1}{\Delta}\left(\begin{array}{ccc}
x_2y_3-x_3y_2&x_3y_1-x_1y_3&x_1y_2-x_2y_1\\
y_2-y_3&y_3-y_1&y_1-y_2\\
x_3-x_2&x_1-x_3&x_2-x_1\\
\end{array}\right)
\end{equation}
where
  $ \Delta=\det({\cal M})=x_1y_2+x_2y_3+x_3y_1-x_1y_3-x_2y_1-x_3y_2 $.
Now one may express the unperturbed state amplitudes, $a_j(t)$,
in terms of the dressed-state amplitudes, $c_j(t)$.  Using
$c_j(t) = e^{-iz_jA(t)}$, one has
$a_i(t) = \sum_{j=1}^3 {\cal M}^{-1}_{ij} c_j(t)
= \sum_{j=1}^3 {\cal M}^{-1}_{ij} e^{-iz_jA(t)}$.
Specifically,
\begin{eqnarray}
\label{aj}
a_1&=&\frac{1}{\Delta}\left((x_2y_3-x_3y_2)e^{-iz_1A(t)}+(x_3y_1-x_1y_3)e^{-iz_2A(t)}
	+(x_1y_2-x_2y_1)e^{-iz_3A(t)}\right)   \nonumber \\
a_2&=&\frac{1}{\Delta}\left((y_2-y_3)e^{-iz_1A(t)}+(y_3-y_1)e^{-iz_2A(t)}
	+(y_1-y_2)e^{-iz_3A(t)}\right) \nonumber \\
a_3&=&\frac{1}{\Delta}\left((x_3-x_2)e^{-iz_1A(t)}+(x_1-x_3)e^{-iz_2A(t)}
	+(x_2-x_1)e^{-iz_3A(t)}\right) \ \ .
\end{eqnarray}
The transition probabilities are given by the analytic expressions,
\begin{eqnarray}
\label{Pj'}
P_1(t)&=&|a_1(t)|^2=\frac{1}{\Delta^2}[(x_2y_3-x_3y_2)^2+(x_3y_1-x_1y_3)^2
	+(x_1y_2-x_2y_1)^2 \nonumber \\
&&+2(x_2y_3-x_3y_2)(x_3y_1-x_1y_3)\cos((z_1-z_2)A(t)) \nonumber \\
&&+2(x_2y_3-x_3y_2)(x_1y_2-x_2y_1)\cos((z_1-z_3)A(t))\nonumber \\
&&+2(x_3y_1-x_1y_3)(x_1y_2-x_2y_1)\cos((z_2-z_3)A(t))] \ \ ,  \nonumber \\
P_2(t)&=&|a_2(t)|^2=
\frac{1}{\Delta^2}[(y_2-y_3)^2+(y_3-y_1)^2+(y_1-y_2)^2\nonumber \\
&&+2(y_2-y_3)(y_3-y_1)\cos((z_1-z_2)A(t))\nonumber \\
&&+2(y_2-y_3)(y_1-y_2)\cos((z_1-z_3)A(t))\nonumber \\
&&+2(y_3-y_1)(y_1-y_2)\cos((z_2-z_3)A(t))] \ \ ,  \nonumber \\
P_3(t)&=&|a_3(t)|^2=
\frac{1}{\Delta^2}
[(x_3-x_2)^2+(x_1-x_3)^2+(x_2-x_1)^2\nonumber \\
&&+2(x_3-x_2)(x_1-x_3)\cos((z_1-z_2)A(t))\nonumber \\
&&+2(x_3-x_2)(x_2-x_1)\cos((z_1-z_3)A(t))\nonumber \\
&&+2(x_1-x_3)(x_2-x_1)\cos((z_2-z_3)A(t))]  \ .
\end{eqnarray}

One may now seek the conditions on the external field $V(t)$ such that the
electron populations $P_j(t) = |a_j(t)|^2$ take desired values.
We have been able to show \cite{rms03} that for $n=3$
the electron can be fully transferred to a targeted state at
an arbitrary time $t = t_0$ by suitably adjusting the $V_{ij}(t)$.

\subsection{$n$-state atoms for $n \geq 4$}

For a $4$-state atom one similarly obtains a 4$^{th}$ order equation in $x$,
yielding four eigenvalues for $z_j$ and four corresponding eigenfunctions
$c_j(t) = e^{-i z_j A(t)}$.  For $n > 4$, analytic solutions to an $n^{th}$
order equation do not exist.  However, one may determine the eigenvalues
$x_j$ by the numerical method of successive approximations \cite{AS}.
It is possible to control the occupation probabilities, $P_k(t)$,
for any $n$ by adjusting the magnitudes of the $V_{ij}$,
e.g. $\alpha$, $\beta$ and $V$ in Eq(\ref{3degen}), and by
changing the shape of $V(t)$.

As will be verified below, there is a general scheme
to find analytic solutions for degenerate $n$-state atoms \cite{msr03}.
This scheme is straightforward.
Seek a solution to Eq(\ref{ndegen}) of the form
$c(t) = x_1 a_1(t) + x_2 a_2(t) + \cdots + x_n a_n(t)$.
Since $a_i(0) = \delta_{i1}$, one has that $x_1 = 1$.  Next calculate
$i \dot{c}(t)$ using Eq(\ref{ndegen}) and require that $i \dot{c}(t) = z V(t) \ c(t)$.
Here $V(t)$ is a common factor for the $V_{jk}(t)$ terms in
Eq(\ref{ndegen}), and $z$ is a linear combination of the $x_i$'s, dependent
on the relative (time independent) strengths of the $V_{jk}(t)$.
This leads to an $n^{th}$ order equation in $x_2$ (or any of the other $x_i$'s
($i \geq 2$)), whose roots may be denoted by $x_{j2}$ (or $x_{ji}$ in general).
This yields $n$ eigenvalues, $z_j$, and $n$ eigenfunctions,
$c_j(t) = e^{- i z_j A(t)}$, where $A(t) = \int_0^t V(t')dt'$.
This process determines the matrix elements, ${\cal M}_{ij}$,
for $c_i(t) = \sum_j^n {\cal M}_{ij} a_j(t)$.  Specifically, ${\cal M}_{ij} = x_{ij}$.
Inverting this relation yields the probability amplitudes for the
electron population, $a_k(t) = \sum_j^n {\cal M}^{-1}_{kj} c_j(t)
= \sum_j^n {\cal M}^{-1}_{kj} e^{-i z_j A(t)}$.
Using $\cos(a-b) = \cos a \cos b + \sin a \sin b$,  one quickly obtains,
\begin{eqnarray}
\label{Pkn}
P_k(t) = |a_k(t)|^2 = \sum_{i}^n \ \sum_{j}^n  \
{\cal M}_{ki}^{-1}{\cal M}_{kj}^{-1}  \cos[(z_i - z_j) A(t)] \ \ .
\end{eqnarray}
Since the $x_{kj}$'s and $z_j$'s vary with the $V_{jk}(t)$,
one may seek conditions on the matrix elements $V_{jk}(t)$
and on $A(t_0)$ such that the electron populations $P_k(t_0) = |a_k(t_0)|^2$
take desired values at $t = t_0$.
It has been shown \cite{rms03,sm02} that complete population transfer
occurs in 2-state and 3-state atoms at $t = t_0$ if $A(t_0)/\pi = 1/2$ in the 2-state atom
and $A(t_0)/\pi = 1 / \sqrt{2}$ in the 3-state atom.  In addition for the 3-state atom
$V_{12}(t)=0$ and $V_{13}(t)=V_{23}(t)$.

We now note that the $n$-state equations simplify if
$V_{ij}(t) = \gamma V_{23}(t)$ for all $j \geq 3$, since
all $a_j(0) = 0$.  Using $V(t) = \alpha^{-1} V_{12}
= \beta^{-1} V_{13} = \gamma^{-1} V_{23} = \epsilon_j^{-1} V_{ii}$,
Eq(\ref{ndegen}) becomes,
\begin{eqnarray}
\label{nsym}
i \dot{a}_1(t) &=& V(t) \ ( \ \epsilon_1 \ a_1(t) + \alpha \ a_2(t) + (n-2) \beta \ a_3(t) \ )
        \ \ , \\ \nonumber
i \dot{a}_2(t) &=& V(t) \ ( \ \alpha \ a_1(t) + \epsilon_2 \ a_2(t) + (n-2) \gamma \ a_3(t) \ )
        \ \ ,  \\ \nonumber
i \dot{a}_3(t) &=& V(t) \ ( \ \beta \ a_1(t)  + \gamma \ a_2(t) + ( \epsilon_3 +  \frac{n-3}{n-2} ) \ a_3 \ )
        \ \ .
\end{eqnarray}
Since $V(t)$ is arbitrary at this point, we may set $\gamma = 1$ without loss of generality.
Note that these equations differs from the equations for true 3-state atoms in that
$V_{ij}(t) \neq V_{ji}(t)$.

These equations may be solved using the general scheme described above.
To find a solution that is mathematically simple, following the
3-state atom \cite{rms03} we choose $\beta = 1$ and $\epsilon_j = \epsilon$.
The value of $\epsilon$ may be arbitrarily changed by
an overall phase transformation of the $a_j$.
Taking $x_2 = x$ and $x_3 = y$ to simplify notation, one quickly obtains,
$x = (\alpha + y)/(\alpha x + y)$ and $y = ( (1 + x) + (\frac{n-3}{n-2}) y )/(\alpha x + y)$.
This yields cubic equations in $x$ and $y$.
However, it is evident in this case that if $x = -1$ then $y = 0$ and
if $x = 1$ then
$y = y_{\pm} = \frac{1}{2}(-\alpha + n - 3 \pm \sqrt{(\alpha + n - 3)^2 + 2/(n-2)}$.
Hence there are three eigenvalues for $x$, $y$ and $z$, namely $\{x_i\} = \{1,1,-1\}$,
$\{y_i\} = \{y_+,y_-,0\}$ and $\{z_i\} = \{y_+ + \alpha,y_- + \alpha, -\alpha\}$.
This gives three eigenfunctions, $c_j = e^{-iz_j A(t)}$, where
$A(t) = \int_0^t V(t') dt'$, which are a linear combination of the $a_i(t)$.
Specifically $a_i = {\cal M}^{-1}_{ij} c_j$, where,
$ {\cal M}^{-1}=\frac{1}{2(y_+ - y_-)}\left(\begin{array}{ccc}
 -y_- & y_+ & (y_+ - y_-)  \\
 -y_- & y_+ & -(y_+ - y_-)  \\
  1  &  -1  &     0         \\
\end{array}\right) $
This yields transition probabilities $P_i(t) = |a_i(t)|^2$, in accord with Eq(\ref{Pkn}).

From Eq(\ref{Pkn}) extrema of $P_i(t)$ occur at $t = t_0$ when
$(z_1 - z_2) A(t_0)/\pi = k$ and $(z_2 - z_3) A(t_0)/\pi = k'$.
The third condition is redundant since $P_1 + P_2 + P_3 = 1$.
One may show after some algebra that these conditions are met
when,
\begin{eqnarray}
\label{qcond}
  A(t_0)/\pi &=& \pm n_0 \sqrt{\frac{9}{18 (n-2) + 4 (\frac{n-3}{n-2})}} \ \ , \nonumber \\
 \alpha &=& V_{12}(t)/V_{23}(t) =  -\frac{1}{3} (n-3)  \ \ , \nonumber \\
 \beta  &=& V_{13}(t)/V_{23}(t) = 1  \ \ ,
\end{eqnarray}
where $n_0$ is any odd integer. One may then show,
\begin{eqnarray}
\label{Pn}
  P_1 &=& |a_1(t)|^2 = \frac{1}{8} [ 3 + \cos(\theta) + 4 \cos(\theta/2) ]  \ \ , \nonumber \\
  P_2 &=& |a_2(t)|^2 = \frac{1}{8} [ 3 + \cos(\theta) - 4 \cos(\theta/2) ]  \ \ , \nonumber \\
  P_3 &=& |a_3(t)|^2 = \frac{1}{2(n-2)} \sin^2(\theta/2) \ \ ,
\end{eqnarray}
where $\theta = \theta(t) = 2 \pi [A(t)/A(t_0)]$  with $A(t) = \int_0^t V(t') dt'$.
Complete population transfer occurs from state 1 to state 2 at $t = t_0$
when $\theta(t) = 2 \pi$.
As $n \to \infty$ these solutions reduce to those of a 2-state atom
with $V_{11} = V_{22}$, where $P_1 = \cos^2[A(t)]$ and $P_2 = \sin^2[A(t)]$.

\section{Results}

\subsection{2-state atom}

In this section we compare full and analytic results for
the $2s-2p$ transition in hydrogen in the two state approximation~\cite{sm02}.

\subsubsection{Population leakage}

To obtain our analytic solution to Eq.(\ref{2a-ans}), we had to neglect the $\omega_{21}a_2$ term in the equation
$ i \dot{a}_2 =  \omega_{21}a_2 - \chi \cos (\omega t) a_1. $
This approximation requires that $\omega_{21} \ll \chi$, or (since  $\chi/\omega=\pi/2$), $\omega_{21} \ll \omega.$
When $\omega_{21}$ is finite, the population transfer is not complete.
The effect of this population leakage can be calculated by expanding the transition amplitude $a_2$ in a power series in time including terms $\omega_{21}$.
It is easily shown that the first $\omega_{21}$ term corresponds to the second derivative, $\ddot{a_2}(t)$. The difference between the exact and the analytic solutions for the transition amplitudes is $ \Delta a_2(t) \approx \frac{1}{2}\omega_{21}\chi t^2 $, and
\begin{equation}\label{delta P}
\Delta P(t) = |\Delta a_2(t)|^2 \approx \frac{1}{4}\omega_{21}^2 \chi^2 t^4 \ .
\end{equation}
By the time the occupation probability $P_2$ reaches its first maximum at $t_0=T/4=\pi/2 \omega$, the difference becomes
\begin{equation}\label{delta P 2}
\Delta P(t_0) \approx \frac{1}{4}(\frac{\pi}{2})^6(\frac{\omega_{21}}{\omega})^2 \ .
\end{equation}

\subsubsection{Calculations of $2s-2p$ transitions in hydrogen}
\begin{figure}
\centering
\scalebox{0.7}{\includegraphics{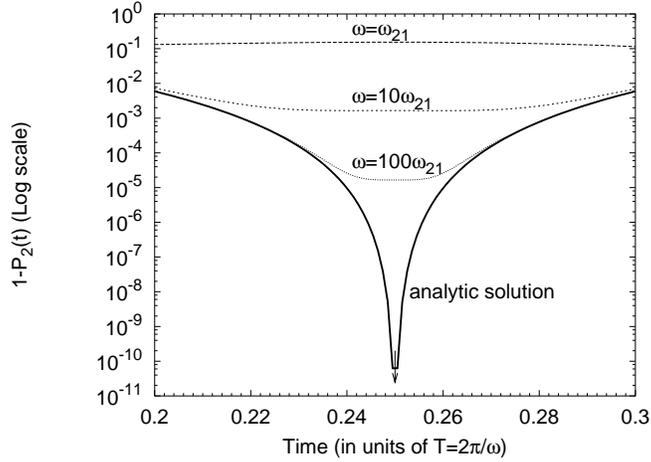}}
\caption{Occupation probability $P_2(t)$ in the vicinity of $t_0=T/4$: analytical solution, Eq.(\ref{f9}), and numerical calculations for different values of the ratio $\omega/\omega_{21}$ (1, 10 and 100). Deviation from the analytic solution decreases as $(\omega_{21}/\omega)^2$ (cf. Eqn. (\ref{delta P 2}))}
\label{fig3}
\end{figure}

Here we compare the full (numerical) calculations and the analytical solution for a $2s-2p$ transition in hydrogen.
The full calculations were done by numerically integrating Eqs. (\ref{2a-ans}) using a standard fourth order Runge-Kutta algorithm.
In hydrogen the energy separation of $2s$ and $2p$ states (Lamb shift) is $4.37*10^{-6}eV$, while the next available state ($3p$) is $\approx 1.89 eV$ away.
Therefore, one can choose the frequency of the external field $\omega$ such that
$\omega_{21} \ll \omega \ll \omega_{2s3p}$, i.e. both the degenerate state approximation (limit $\omega_{21} \rightarrow 0$) and the two-state model can be used.

As shown in figure \ref{fig3}, the difference between the full and analytic solutions for $P_1(t)=1-P_2(t)$ in the vicinity of $t_0=\pi/2\omega$  is large when $\omega=\omega_{21}$.
But it decreases for $\omega > \omega_{21}$.
The difference is less than 1\% for $\omega/\omega_{21}=10$, and  0.01\% for $\omega/\omega_{21}=100$.
Since that difference is proportional to  $(\omega_{21}/\omega)^2$, one can go up in frequency as high as fractions of $eV$ (that brings the difference between the approximate and exact solution down to $10^{-10}-10^{-11}$) and yet be far sensibly from the nearest available ($2s-3p$) resonant transition  frequency.
Therefore, a  radiation field with the wavelength from a few $\mu m$ (and the intensity of the order $10^{12} W/cm^2$) up to a few $cm$ (and the intensity of the order $10^4 W/cm^2$) can be used for $2s-2p$ transitions in hydrogen.
The probability of the multiphoton excitation to the $3p$ state (or any other state, including continuum) also seems to be small for the range of the external field frequencies under consideration.

Our analytic approximation appears to be valid
for $2s-2p$ transitions in hydrogen over a broad range of
field frequencies.

\subsubsection{Changing the shape of $V(t)$}\label{shape}
We have noted previously that the choice of the field frequency is a trade-off between two competing factors: reducing the population leakage and increasing the duration of the populated  state.
If wishes to obtain a long lasting populated state with very small leakage, it may be that
both requirements for the system cannot be met simultaneously for a single frequency radiation field.
In this case it may be possible to use another form of the interaction potential.
In particular, one may change the shape of the "flat" part of the probability (cf. figure \ref{fig1}) by using different shapes for the external potential $V_{ext}(t)$ . For an arbitrary external potential the formulas for the occupation probabilities can be written in the form:
\begin{eqnarray}
\label{a4eps}
&& P_1(t) = \cos^2 \lbrack \int_0^t V_{21}(\tau)d\tau \rbrack \nonumber\\
&& P_2(t) = \sin^2 \lbrack \int_0^t V_{21}(\tau)d\tau \rbrack \ .
\end{eqnarray}

A Taylor series expansion can be used for choosing the shape of the external potential. For example, one can use the potential with the first non-vanishing derivative of the order higher than four to make the shape of the populated state even flatter.
The derivatives can be calculated using the general formula for the n$th$ derivative of a composite function \cite{Gradshteyn}, which in this case takes the form
\begin{equation}\label{gradshteyn}
\frac{d^n P_2}{d t^n}= \sum \frac{n!}{i! j! \dots k!} \frac{d^m F}{d y^m}
(\frac{y^{\prime}}{1!})^i (\frac{y^{\prime \prime}}{2!})^j \dots (\frac{y^{(l)}}{l!})^k \ ,
\end{equation}
where $P_2(t)=F(y)=\sin^2 y \ $, $ \ y=y(t)=\int_0^t V_{21}(\tau) \ d\tau \ $, $ \ \sum$ indicates summation over all solutions in non-negative integers of the equation $i+2j+\dots +lk=n$ and $m=i+j+ \dots +k$.

In principle, Eqn.(\ref{gradshteyn}) tells us how to shape $V(t)$ to control the population transfer.
As one can see from Eq.(\ref{gradshteyn}), one may eliminate all terms up to the order $\tau^k$ by choosing the potential for which all derivatives up to the order $(k-1)$ are zero at the point $t=t_0$.
That will "flatten" the shape of the populated (or depopulated) state, i.e. allow one to use smaller frequency (and, as a result, an increased duration of the state) to achieve the same state of population leakage.
Therefore, the shape of an external potential can be used (along with the choice of the field frequency) for quantum control.
For example, in a truly two-state system, choosing the potential of the form
$$V(t)=\frac{\pi}{2}\delta(t-t_0)$$
leads to
\begin{eqnarray}\label{f10}
&& P_1(t)=1-\Theta(t-t_0) \nonumber\\
&& P_2(t)=\Theta(t-t_0) \ .
\end{eqnarray}
This represents complete and immediate population inversion at $t=t_0$.
However, Fourier transformation of $V(t)$ now contains all frequencies, so one can no longer use a two-state approximation for the $2s-2p$ transition in hydrogen, since higher states become necessarily involved.

\subsection{3-state atom}

Some allowed values of the action integral, $A(t_0)$, and the relative
interaction strength, $\alpha$, are given in table I.
This leads to certain allowed values of the action integral $A(t_0)$ and the relative
interaction strength, $\alpha$ (see \cite{rms03}), namely,
\begin{eqnarray}
\label{qnumbers'}
& \pm 3 \sqrt{\frac{2}{n_1n_2}} \ \ A(t_0) = \pi  \ \ , \nonumber \\
&\alpha = V_{12}(t)/V_{23}(t) = \pm \sqrt{\frac{2}{n_1n_2}} \ (n_1 - n_2)  \ \ , \nonumber \\
&\beta  = V_{13}(t)/V_{23}(t) = \pm 1  \ \ .
\end{eqnarray}

This table includes values of $\{n_1,n_2\}$, $\{n_o,n_o'\}$ and
all three $\{k,k'\}$ cases defined in~\cite{rms03}.
From more complete numerical output we confirm that $n_1$ and $n_2$ each
acquire all possible odd integer values, although values of
the product $n_1 \cdot n_2$ are restricted.
This is consistent with the condition that
$n_1 = 2 n_o + n_o'$, and $n_2 = n_o + 2 n_o'$,
where $n_o$ and $n_o'$ are arbitrary odd integers.
It is also evident that the even integer, $n_e = n_o + n_o'$,
takes on all even values.
One may also show algebraically that for each value of
the even integer $k$ in case i) there are two odd values of
an odd integer $k$ in case ii), and vice versa.
The sets of integers $\{n_1,n_2\}$, $\{n_o,n_o'\}$ and $\{k,k'\}$ are redundant.
The three sets of $\{k,k'\}$ correspond to a single set $\{n_1,n_2\}$,
while the $\{n_o,n_o'\}$ are in one to one correspondence with the $\{n_1,n_2\}$.

\begin{table}
\centering
\label{tab:1}
\begin{tabular}{||c|cc|cc|cc|cc|c|c||}
\hline
\multicolumn{1}{||c|}{$n_1\cdot n_2$} &\multicolumn{1}{c}{$n_1$} &\multicolumn{1}{c|}{$n_2$}
&\multicolumn{1}{c}{$n_e$} &\multicolumn{1}{c|}{$n_o'$}
&\multicolumn{1}{c}{$n_o$} &\multicolumn{1}{c|}{$n_o'$}
&\multicolumn{1}{c}{$n_o$} &\multicolumn{1}{c|}{$n_e$}
&\multicolumn{1}{c|}{$A(t_0)$} &\multicolumn{1}{c||}{$\alpha$} \\
& & & $k$ & $-k'$ & $k$ & $k'$ & $-k$ & $k'$ & & \\
\hline
5 \ \ \  &$\pm 1$  &$\pm 5$  &$\pm 2$ &$\pm  3$ &$\mp  1$ &$\pm  3$ &$\mp  1$ &$\pm 2$ &$\pm 1.656$ &$\mp 2.530$ \\
5 \ \ \  &$\pm 5$  &$\pm 1$  &$\pm 2$ &$\mp  1$ &$\pm  3$ &$\mp  1$ &$\pm  3$ &$\pm 2$ &$\pm 1.656$ &$\pm 2.530$ \\
9 \ \ \  &$\pm 3$  &$\pm 3$  &$\pm 2$ &$\pm  1$ &$\pm  1$ &$\pm  1$ &$\pm  1$ &$\pm 2$ &$\pm 2.221$ &$    0.000$ \\
11\ \ \  &$\pm 1$  &$\pm 11$ &$\pm 4$ &$\pm  7$ &$\mp  3$ &$\pm  7$ &$\mp  3$ &$\pm 4$ &$\pm 2.456$ &$\mp 4.264$ \\
11\ \ \  &$\pm 11$ &$\pm 1$  &$\pm 4$ &$\mp  3$ &$\pm  7$ &$\mp  3$ &$\pm  7$ &$\pm 4$ &$\pm 2.456$ &$\pm 4.264$ \\
17\ \ \  &$\pm 1$  &$\pm 17$ &$\pm 6$ &$\pm 11$ &$\mp  5$ &$\pm 11$ &$\mp  5$ &$\pm 6$ &$\pm 3.053$ &$\mp 5.488$ \\
17\ \ \  &$\pm 17$ &$\pm 1$  &$\pm 6$ &$\mp  5$ &$\pm 11$ &$\mp  5$ &$\pm 11$ &$\pm 6$ &$\pm 3.053$ &$\pm 5.488$ \\
23\ \ \  &$\pm 1$  &$\pm 23$ &$\pm 8$ &$\pm 15$ &$\mp  7$ &$\pm 15$ &$\mp  7$ &$\pm 8$ &$\pm 3.551$ &$\mp 6.487$ \\
23\ \ \  &$\pm 23$ &$\pm 1$  &$\pm 8$ &$\mp  7$ &$\pm 15$ &$\mp  7$ &$\pm 15$ &$\pm 8$ &$\pm 3.551$ &$\pm 6.487$ \\
27\ \ \  &$\pm 3$  &$\pm 9$  &$\pm 4$ &$\pm  5$ &$\mp  1$ &$\pm  5$ &$\mp  1$ &$\pm 4$ &$\pm 3.848$ &$\mp 1.633$ \\
27\ \ \  &$\pm 9$  &$\pm 3$  &$\pm 4$ &$\mp  1$ &$\pm  5$ &$\mp  1$ &$\pm  5$ &$\pm 4$ &$\pm 3.848$ &$\pm 1.633$ \\
29\ \ \  &$\pm 1$  &$\pm 29$ &$\pm10$ &$\pm 19$ &$\mp  9$ &$\pm 19$ &$\mp  9$ &$\pm10$ &$\pm 3.988$ &$\mp 7.353$ \\
29\ \ \  &$\pm 29$ &$\pm 1$  &$\pm10$ &$\mp  9$ &$\pm 19$ &$\mp  9$ &$\pm 19$ &$\pm10$ &$\pm 3.988$ &$\pm 7.353$ \\
35\ \ \  &$\pm 1$  &$\pm 35$ &$\pm12$ &$\pm 23$ &$\mp 11$ &$\pm 23$ &$\mp 11$ &$\pm12$ &$\pm 4.381$ &$\mp 8.128$ \\
35\ \ \  &$\pm 5$  &$\pm 7$  &$\pm 4$ &$\pm  3$ &$\pm  1$ &$\pm  3$ &$\pm  1$ &$\pm 4$ &$\pm 4.381$ &$\mp 0.478$ \\
35\ \ \  &$\pm 7$  &$\pm 5$  &$\pm 4$ &$\pm  1$ &$\pm  3$ &$\pm  1$ &$\pm  3$ &$\pm 4$ &$\pm 4.381$ &$\pm 0.478$ \\
35\ \ \  &$\pm 35$ &$\pm 1$  &$\pm12$ &$\mp 11$ &$\pm 23$ &$\mp 11$ &$\pm 23$ &$\pm12$ &$\pm 4.381$ &$\pm 8.128$ \\
\hline
\end{tabular}
\caption{Some values for the action integral, $A(t_0)$, and the relative
interaction strength, $\alpha$, allowed for total population transfer.
These values are found using Eq(\ref{qnumbers'}) subject to the conditions listed.
Here $n_e = n_o + n_o'$, $n_o = \frac{1}{3}(2n_1 - n_2)$
and $n_o' = \frac{1}{3}(2n_2 - n_1)$.
}
\end{table}

\begin{figure}
\centering
\scalebox{0.7}{\includegraphics{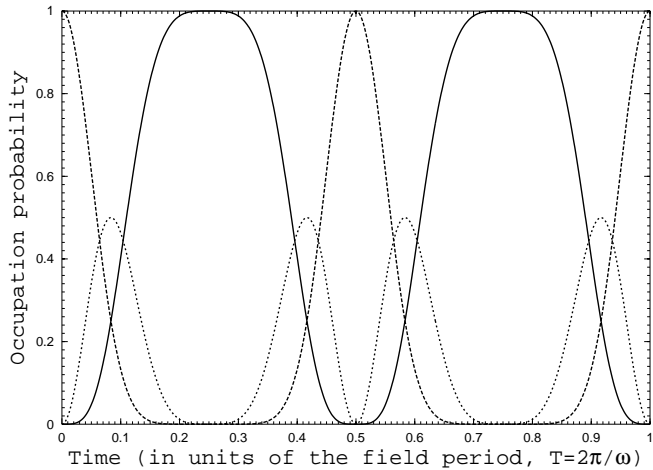}}
\caption{Occupation probabilities as a function of time, Eq. (\ref{Pj'}).
The solid line denotes $P_2(t)$; the long dash line denotes $P_1(t)$;
and the short dash line denotes $P_3(t)$.
In this figure we use the  allowed values, $\alpha = V_{12}/V_{23} = 0$ and
$A(t_0) = \int_0^{t_0} V(t') dt' = 2.221$, corresponding
to $n_1 = 3$ and $n_2 = 3$ ($n_o = n_o' = 1$) in Eq(\ref{qnumbers'}).
Complete transfer to state 2 from state 1 occurs at $t = t_0 = T/4$
and again at odd multiples of $t_0$.
}
\label{fig4}
\end{figure}

\begin{figure}
\centering
\scalebox{0.7}{\includegraphics{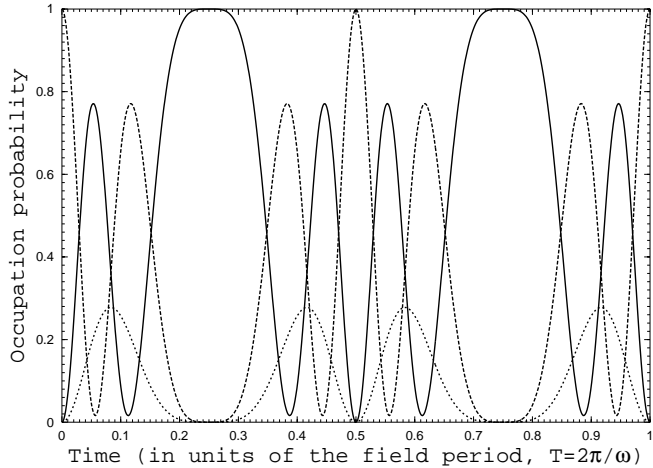}}
\caption{Occupation probabilities as a function of time.
The solid line denotes $P_2(t)$; the long dash line denotes $P_1(t)$;
and short dash line denotes $P_3(t)$.
In this figure $\alpha = -2.530$ and $A(t_0) = 1.656$, corresponding
to $n_1 = 1$ and $n_2 = 5$ ($n_o= -1$, $n_o' = 3$) in Eq(\ref{qnumbers'}).
}
\label{fig5}
\end{figure}

\begin{figure}
\centering
\scalebox{0.7}{\includegraphics{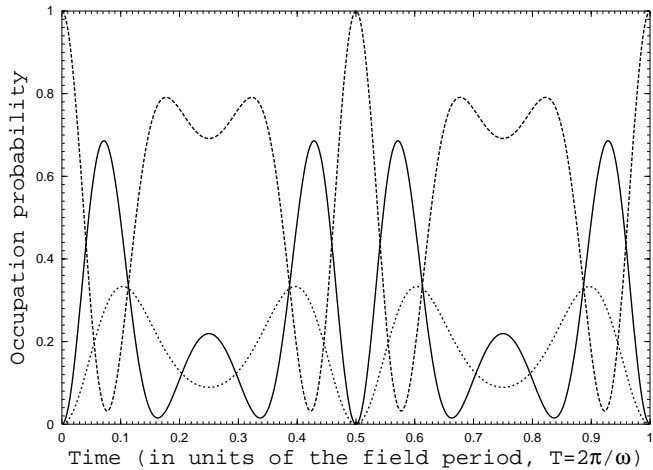}}
\caption{Occupation probabilities as a function of time.
The solid line shows $P_2(t)$; the long dash line shows $P_1(t)$;
and the short dash line shows $P_3(t)$.
Here $\alpha = 8.128$ and $A(t_0) = 4.381$, corresponding
to $n_1 = 35$ and $n_2 = 1$ ($n_o= 23$, $n_o' = -11$) in Eq(\ref{qnumbers'}).
Since these values are allowed, the electron population
is completely transferred to state 2 from state 1.
}
\label{fig6}
\end{figure}

Numerical calculations for the time dependence of the populations
in a degenerate $3$-state atom perturbed by external interactions with
$V(t) = V_0 \cos(\omega t)$ are presented in figures~\ref{fig4}-\ref{fig6}.
These results were obtained by using a standard fourth order
Runge-Kutta numerical integration of Eq(\ref{3degen}) with $\epsilon_j =0$
and $\beta = 1$ for various values of $\alpha$ and $A(t_0)$.
Again complete transfer to an initially
unoccupied state never occured.  However, there were rapid oscillations
in the populations of all states except near $t = T/4$ and $t = 3T/4$,
where none of the populations oscillated rapidly.  This was similar to figure~\ref{fig6}.
This appears to correspond to the onset of complete population transfer,
which occurs at $t = T/4$ and odd multiples of $T/4$, as seen in
figures~\ref{fig4} and~\ref{fig5}.

Calculations using a few values of $\alpha$ that permit complete transfer
to state 2 are shown in figures~\ref{fig4} and~\ref{fig5}.  Our numerical codes give the same
results as the analytic expressions of Eq(\ref{Pj'}).
For $n$-state systems Eq(\ref{Pn}) gives population transfers from state 1 to state $n$.
This is shown in figure~\ref{fig4}, where the short dash line corresponds to sum of $P_3$, $P_4$, $\cdots$, $P_n$.
This provides a check that our algebra is correct.
We note that $\frac{2 n_1 n_2}{(n_1 + n_2)^2} \leq \frac{1}{2}$ and
that the maximum value of $P_3(t)$ occurs for $n_1 = n_2$, where
$P_{3 \ max} = \frac{1}{2}$, consistent with figure~\ref{fig4}.  This corresponds
to $\alpha =0$ so that direct transitions from state 1 to state 2 are
forbidden.  Transfer to state 2 occurs via the intermediate state 3.
Transfer from state 1 to state 2 and back is complete, and
occurs periodically.
In general $\alpha = 0$ corresponds to $n_1 = n_2 = 3n_{odd}$,
where $n_{odd}$ is any odd integer.
This appears to give the simplest condition that allows complete
population transfer.
In this case the action area is $A(t_0) = n_{odd} \pi / \sqrt{2}$.

Calculations for two other values of $\alpha$ that allow complete transfer
to state 2 are shown in figures~\ref{fig4} and~\ref{fig5}.  We see that
complete transfer occurs twice in one period of the oscillating field
but that the frequency of the "side bands" increases as $\alpha A$ increases.
We note that $\alpha = \pm \sqrt{\frac{2}{n_1 n_2}} \ (n_1 - n_2)$ becomes either large
($n_1 \gg n_2$, or vice versa), or small ($n_1 \sim n_2$) as $n_1 n_2$ increases,
while $A(t_0)$ increases as $\sqrt{n_1 n_2/2}$.
When complete population transfer occurs, the population lingers in state 2,
as seen in figures~\ref{fig4} and~\ref{fig5}.
It can be shown \cite{sm02} near $t = t_0$ that $1 - P_2(t)$
varies as $[\omega (t - t_0)]^4$.

Additional control \cite{sm02} may be achieved by changing the shape of $V(t)$.
This can be used to control how long the population remains near unity
in state 2, for example.  An interesting example
is the case of an ideal, sudden 'kick' produced by $V(t) = A_0 \delta(t - t_0)$.
If $A_0 \to n_{odd} \pi / \sqrt{2}$ for example, then the mere
presence of state 3 allows transfer from state 1 to state 2
without any direct transfer from state 1 to state 2.
With this ideal 'kick', state 2 is unoccupied before $t = t_0$,
and fully occupied after $t = t_0$.

To obtain analytic solutions to Eq.(\ref{ndegen}) for a 3-state atom,
we imposed degeneracy on the 3-state manifold, e.g. by taking $E_j = E = 0$.
If $E \neq 0$, then $E$ can be removed from Eq(\ref{nexact})
by an overall phase transformation, $e^{i E t}$.
When the states are not degenerate $\omega_{ij} = (E_i - E_j)/\hbar \neq 0$, and
the population transfer is incomplete.
Numerical calculations indicate that the population leakage
varies as $(\omega_{ij}/\omega)^2$.
The choice of the field frequency, $\omega$, involves a trade-off between the duration
of time the population remains in state 2 and the population leakage.
The higher the frequency, $\omega$, the smaller the population leakage, $\Delta P_2$,
but the shorter the duration time, $T_s$, that the population remains in state 2.
This effect of population leakage is similar as that for a 2-state atom \cite{sm02}.
Numerical calculations indicate that this population leakage grows rapidly in time.

\section{Discussion}

In this work we have used the $2s-2p$ transition in hydrogen as an example of population transfer.
The same approach can be used for any other atomic or molecular system that has a similar pattern of energy states (two states located close one to another and far from the other states).
As we have shown in Eq.(\ref{delta P 2}), for an external field with a single frequency $\omega$, the difference between the exact calculations and the analytic approximation varies as $(\omega_{21}/\omega)^2$ for $\omega_{21}/\omega \ll 1$, and does not depend on the internal structure of the atom or molecule.
This feature opens the possibility for using different systems with different values of the transition frequency, $\omega_{21}$ and different ranges for the field frequency, $\omega$.

A downside of analyzing more complex atomic or molecular systems is that analytic expressions for the orbital functions are not always available.
Then genetic algorithms (GA) may be used to choose a shape for the external potential, $V(t)$.
The application of GA with active feedback to the selective breaking and making chemical bonds in polyatomic molecules has been discussed in detail by Rabitz {\it et al } \cite{Rabitz93, Rabitz92, Rabitz2000}, and tested experimentally \cite{Levis99, Levis01}).
As in any optimization scheme, the effectiveness of the GA increases significantly when the initial value of the parameter close to the optimal one is chosen.
One might combine our method with GA using approximate analytic orbital functions to choose the "starting" form of the potential, estimate $T_s$ and $P_{cr}$, and then employ the GA scheme.

For $2s-2p$ transitions in hydrogen, sources of the microwave radiation with corresponding intensities may now be available, which can be used to test our model.
Metastable excited $H(2s)$ atoms could interact with a low frequency (e.g. microwave) radiation.
Since the lifetime of the $2p$ state is much shorter than the $2s$ state, the population of the $2p$ state may be monitored by observing photons emitted in $2p-1s$ transitions.
In these experiments the smallness of the Lamb shift requires the use of the temperatures below the $1 \ mK$  to exclude the thermal transitions between the states.

In this work we have also considered transitions amplitudes and
probabilities in a $3$-state atom with degenerate states.
We also assumed that the interaction matrix elements, $V_{ij}(t)$
all have a common time dependence.  In realistic atomic
systems, energy states are seldom, if ever, exactly degenerate.
Also states outside $3$-state manifold usually exist.
Consequently, use of our results is restricted to external fields
with frequencies in the range, $\omega_{min} < \omega <  \omega_{max}$.
Here $\hbar \omega_{min}$ is the energy splitting of the nearly
degenerate states, and $\hbar \omega_{max}$ is the energy difference
between the $3$-state manifold and the closest state in energy outside
the manifold.

It is instructive to compare our 3-state results to
simpler 2-state results \cite{msr03,sm02}.
The equations for a $2$-state atom may be recovered from our $3$-state
equations by taking $V_{12}$ and $V_{13}$ to zero, which corresponds
to $\alpha \to \infty$ (with any $|\epsilon_j/\alpha| < \infty$),
as may be understood from Eq(\ref{3trial}).
For the $2$-state atom it has been
shown \cite{msr03,sm02} that $a_j(t)$ may be expressed
in terms of eigenstates $c_i(t)$ via the relation,
$c_i(t) = \sum_{j=1}^2 {\cal M}_{ij} a_j(t)$, where
${\cal M}_{ij}=\left(\begin{array}{cc}
1&y_+\\
1&y_-\\
\end{array}\right)$.
This matrix may be inverted to give the population amplitudes,
$a_1(t) = \frac{1}{\Delta}[y_- c_1(t) - y_+ c_2(t)] $
and $a_2(t) = \frac{1}{\Delta}[ - c_1(t) +  c_2(t)]$,
where $\Delta = 2 \sqrt{1 + (\frac{\epsilon_2 - \epsilon_1}{2})^2}$
is the determinant of ${\cal M}_{ij}$.
A simple solution occurs when $\epsilon_1 = \epsilon_2 = \epsilon$,
namely,
$ a_1(t) = e^{-i \epsilon A(t)} \ \frac{1}{2} [e^{-iA(t)} + e^{+iA(t)}]
	=  \ e^{-i \epsilon A(t)} \cos(A(t))$ and
$ a_2(t) = e^{-i \epsilon A(t)} \ \frac{1}{2} [e^{-iA(t)} - e^{+iA(t)}]
	= - i e^{-i \epsilon A(t)} \sin(A(t)) $.
Note that $P_1 = 1 - P_2$.
One may then simply determine the conditions under which the population
of state 2, $P_2(t_0)$, takes on any desired value at time $t_0$.
If $\epsilon_1 \neq \epsilon_2$, it is easily shown that
$P_2(t_0) \leq 1/(1+ (\epsilon_2 - \epsilon_1)^2/4 ) < 1$.
Thus the diagonal matrix elements of $V_{ij}(t)$ prevent complete transfer
to the initially unoccupied state when $\epsilon_1$ and $\epsilon_2$ are
unequal, and, when they are equal, they simply contribute an overall phase.
When $\epsilon_1 = \epsilon_2$ any value of $P_2(t_0)$ between 0 and 1 can be found.
In particular, $P_2(t_0)=1$ if
$A(t_0) = \int_0^{t_0} V_{12}(t') dt' = n_{odd} \pi /2$.
This allowed value of $A(t_0)$ for complete transfer to state 2 differs
from the allowed values given in Eq(\ref{qnumbers'}) for degenerate 3-state atoms.
By comparison to a 2-state atom the conditions for complete transfer
are generally more complex in a $3$-state atom, as discussed in~\cite{rms03}.
For example, complete population transfer can occur in a 3-state atom
when $\epsilon_2 \neq \epsilon_3$.
In both cases, however, the duration of time spent in the transferred state
can be controlled by adjusting the time dependence of the external field.
Also if the states are not quite degenerate then population leakage occurs.
For a harmonic $V(t)$ the population leakage in a 2-state atom varies as
$\Delta P_2(t_0) \simeq \frac{1}{4} (\frac{\pi}{2})^6 (\omega_{12}/\omega)^2$.
For $2s-2p$ transitions in atomic
hydrogen complete population control can be nearly achieved
in this manner using a radiation field with wavelengths (and intensities)
ranging from a few $\mu m$ (with about $10^{12}$ W/cm$^2$)
up to a few $cm$ (with about $10^4$  W/cm$^2$).
At wavelengths below a few $\mu m$ coupling to nearby $n = 3$ atomic states
can be significant, and this $2$-state model breaks down.

While we have not provided calculations in this work for specific experiments,
some general guidelines for experimental tests and applications are evident.
First, the model must be valid, so that the frequency of the external interaction
is limited by $\omega_{min} < \omega < \omega_{max}$.
If the external interaction varies harmonically, $V(t) = V_0 \cos(\omega t)$,
then the first line of Eq(\ref{qnumbers'}) imposes a constraint between
$V_0$ and $\omega$, namely, $V_0/\omega = \pm \sqrt{\frac{n_1 n_2}{2}} \frac{\pi}{3}$.
For a harmonic interaction the duration of the time, $T_s$, that the population remains
in state 2 varies inversely with $\omega$;  the higher the frequency,
$\omega$, the smaller the time the population remains in state 2.
Our degenerate 3-state model can be applied to systems with dipole
selection rules.  A dipole selection rule can correspond to $\alpha = 0$.
Transfer from state 1 to state 2 (which is the dipole forbidden transition)
is complete at time $t_0$ if the two allowed transition matrix elements,
$V_{23}$ and $V_{13}$, are equal in absolute magnitude and $A(t_0) = n_{odd} \pi/2$.
In some atomic systems state 2 could decay, with a lifetime, $T_d$, to another
state outside the degenerate manifold and be lost.
Such loss can be controlled by adjusting $\omega$.
Our model also rests on degeneracy.
If the three states are not degenerate the transfer of population to
state 2 is incomplete.  We call this population leakage.
As discussed above this population leakage may be minimized
by using high frequency external interactions. i.e. $\omega \gg \omega_{min}$,
but at a cost to duration of time, $T_s$, the population remains in state 2.

Using nearly degenerate states, the duration time, $T_s$,
can be further controlled \cite{sm02} by adjusting the shape of $V(t)$.
As mentioned in section III for example, the transfer is complete, instantaneous
and permanent when $V(t) = n_{odd} \frac{\pi} {\sqrt{2}}  \delta(t - t_0)$.
Such a quick, hard pulse is called \cite{dunning99,burg02,dunning02}
a 'kick'.  Two practical limitations on this ideal model are
the impossibility of producing a signal that varies as $\delta(t - t_0)$,
and the existence of an infinitely wide spectrum of high frequency components
with frequencies $\omega > \omega_{max}$ in the Fourier spectrum of $\delta(t - t_0)$.
Fortunately these two difficulties can both be addressed by using
'kicks' of finite width in time.
In some cases it may be possible to design a 'kick'
so that its duration is short compared to any other changes in the system,
so that a finite 'kick' may be sensibly represented by $\delta(t - t_0)$.
If, in addition, the energy states outside the (nearly) degenerate manifold
have a large energy gap $\hbar \omega_{max}$, then it may be possible
that $\omega_{min} < \omega < \omega_{max}$ and our model may be applicable.
Applying $V(t) \simeq \delta(t - t_0)$ to nearly degenerate atomic systems
leads to the 'gedanken' question, what happens when one tries
to force the transition to occur within a small time interval about $t_0$
in a degenerate quantum system where $\Delta t$ is large?

We note that at sufficiently high $\omega$ all bound states in the atomic elements
become nearly degenerate.
Hence, if one can deal with high energy continuum states,
our approach might be useful in applications involving fourth generation synchrotrons
that produce intense high frequency fields.
An extension of this method might also used for control transitions in high Rydberg states
\cite{Gallagher}, including adiabatic rapid passage \cite{hk83,gallagher02}.
RWA is used to describe coherent storage of information in photonic states \cite{mair02}.
The approach developed here may be useful for modeling information
transmission and storage in atomic states \cite{cirac97} in a new way.

Features that occur when degenerate states are used and the connection to quantum time ordering are discussed elsewhere in this book~\cite{paper1}.

\section{Summary}

We have shown that in a $n$-state atom with degenerate
energies, electron population is completely transferred via
an external interaction at a designated time, $t_0$,
as example we show for $n=2$ and $n=3$.
A new analytical solution of two coupled channel equations
has been found that enables one to temporally control the
electron population of $2s$ and $2p$ states in hydrogen by
using a time varying external field.  In addition, the
population leakage can both be easily estimated, and than
be further controlled by changing the shape of the external
field.
For 3-state atoms from an initially occupied state (state 1) to a designated
initially unoccupied state (state 2) under two conditions.
The first condition for complete transfer is that the ratio
of the matrix elements of the external interaction,
$V_{ij}(t)$,satisfy $V_{12}(t)/V_{23}(t) = \alpha = \pm
\sqrt{\frac{2}{n_1 n_2}} (n_1 - n_2)$, and
$V_{13}(t)/V_{23}(t) = \beta = \pm 1$.  The second
condition is that at $t = t_0$ the action area of $V(t)$
satisfy $A(t_0) = \int_0^{t_0} V(t') dt' = \pm
\sqrt{\frac{n_1 n_2}{2}} \frac{\pi}{3}$, where we have set
$V_{23}(t) = V(t)$.  Here $n_1$ and $n_2$ are integers such
that $n_1 = 2 n_o + n_o'$ and $n_2 = n_o + 2n_o'$, where
$n_o$ and $n_o'$ are any odd integers.  The duration of
time the transferred population remains in state 2 can be
controlled either by varying the frequency, $\omega$, of
the external potential, or by varying the shape of $V(t)$.

This work was supported by the Division of Chemical Sciences,
Office of Science, U.S. Department of Energy.
KhR is supported by a NSF-NATO Fellowship.


\end{document}